\documentclass{JHEP3} 

\usepackage{epsfig,multicol,bbm}

\newcommand\fverb{\setbox\fverbbox=\hbox\bgroup\verb}
\newcommand\fverbdo{\egroup\medskip\noindent%
            \fbox{\unhbox\fverbbox}\ }
\newcommand\fverbit{\egroup\item[\fbox{\unhbox\fverbbox}]}
\newbox\fverbbox

\newcommand{\be}{\begin{eqnarray}}
\newcommand{\ee}{\end{eqnarray}}

\def\ben{\begin{equation}}
\def\een{\end{equation}}
\def\bena{\begin{eqnarray}}
\def\eena{\end{eqnarray}}

\title{Cosmological Perturbations in the New Higgs Inflation}

\author{Cristiano Germani\\
    Arnold Sommerfeld Center, Ludwig-Maximilians-University, Theresienstr. 37\\ 80333 Muenchen, Germany\\
    E-mail: \email{cristiano.germani@lmu.de}}
\author{Alex Kehagias\\
    Physics Division, National Technical University of Athens,\\ 15780 Zografou Campus,  Athens, Greece\\
    E-mail: \email{kehagias@central.ntua.gr}}

\abstract{We study the cosmological perturbations created during the New Higgs inflationary phase. In the New Higgs Inflation, the Higgs boson is kinetically coupled to the Einstein tensor and only three perturbative degrees of freedom, a scalar and two tensorial (gravitational waves), propagate during Inflation.
Scalar perturbations are found to match the latest WMAP-7yrs data within Standard Model Higgs parameters. Primordial gravitational waves also, although propagating with superluminal speed, are consistent with present data. Finally, we estimate the values of the parameter of the New Higgs Inflation in relation to the Higgs mass, the spectral index and amplitude of the primordial scalar perturbations showing that the unitarity bound of the theory is {\it not} violated.
 }

\begin{document}


\section{Introduction}

The latest cosmological data \cite{wmap}  agree impressively well with the assumption that our Universe is, at large scales, homogeneous, isotropic and spatially flat, {\it i.e.}, that it is well described by a Friedmann-Robertson-Walker (FRW) spatially flat geometry. This observation is however a theoretical puzzle. A flat FRW Universe is in fact an extremely fine tuned solution of Einstein equations \cite{liddle}. In the last twenty years or so many attempts have been put forward to solve this puzzle (see for example \cite{many}).
However, the most developed and yet simple idea still remains Inflation \cite{guth}. Inflation solves homogeneity, isotropy and flatness problems in one go just by postulating a
rapid expansion of the early time Universe post Big Bang.

A phenomenological way to achieve Inflation has been pioneered by considering a ``slow rolling'' scalar field \cite{chaotic} with canonical or even
non-canonical kinetic term \cite{k} and lately by non-minimally coupled p-forms \cite{pnflation,proceeding}.
Nevertheless, a fundamental realization of Inflation is still eluding us.

The most economical and yet fundamental candidate for the Inflaton is the Standard Model Higgs boson.
Unfortunately though, the Standard Model parameters are such that no ``slow rolling``
Inflation is possible with the Higgs boson, if minimally coupled to gravity \cite{chaotic}.
However, it has been shown in \cite{higgsinf}, that a non-minimal coupling of the Higgs kinetic term with the Einstein tensor reproduces a successful Inflation without the drawbacks of earlier attempts \cite{sha} \footnote{The inconsistency of the theory \cite{sha} is yet under debate \cite{dub2}. The arguments of \cite{dub} seems to lead to the conclusion that the model of \cite{sha} violates the unitarity bound during Inflation. However, the cut-off of the effective theory of \cite{sha} found in \cite{dub} is too strong, as based on calculations about the wrong background (Minkowski). If instead the Inflationary background is used, the cut-off of the theory is $\it barely$ not exceeded (see also \cite{dub3}). Nevertheless, one should check whether UV corrections to the theory \cite{sha}, will not spoil the perturbation spectrum of the theory \cite{sha}, {\it i.e.} its predictability.}.

In this paper we will consider the primordial, scalar and tensorial, perturbations of the New Higgs Inflation of \cite{higgsinf} and show that the New Higgs Inflation is compatible with observations.

\section{New Higgs Inflation}

The tree-level action of the New Higgs Inflation is
\be
S=\int d^4x\sqrt{-g} \left[\frac{R}{2\kappa^2}-\left(g^{\mu\nu}-w^2
G^{\mu\nu}\right){\cal D}_\mu{\cal H}^\dag {\cal D}_\nu{\cal H}-\lambda\left({\cal H}^\dag{\cal H}-v^2\right)^2\right]\ ,
\ee
where $R$ is the Ricci scalar, $G^{\mu\nu}=R^{\mu\nu}-\frac{R}{2}g^{\mu\nu}$ the Einstein tensor, $\kappa$ the gravitational coupling, $w$ an inverse mass parameter,
${\cal H}$ the  Higgs boson doublet,
${\cal D}_\mu$ the covariant derivative with respect to $SU(2)\times U(1)$
and finally $v$ is the vev of the Higgs in the broken phase of the Standard Model.
In the spirit of chaotic Inflation \cite{chaotic}, we will assume that during Inflation no
 interactions with gauge fields are turned on. With these assumptions we can work with the simpler action
\be
S=\int d^4x\sqrt{-g} \left[\frac{R}{2\kappa^2}-\frac{1}{2}\left(g^{\mu\nu}-w^2
G^{\mu\nu}\right)\partial_\mu\Phi \partial_\nu\Phi-\frac{\lambda}{4}\Phi^4\right]\ ,\label{theory}
\ee
where the Higgs has been rotated as                                    
${\cal{H}}^T=(0,\frac{v+\Phi}{\sqrt{2}})$, and the real scalar $\Phi$ is assumed to be much grater than $v$ during Inflation. We will show later on that this assumption is consistent.

To study a FRW solution of this system, we can directly insert into the action the following metric ansatz
\be
ds^2=-N(t)^2dt^2+a(t)^2\delta_{ij}dx^i dx^j\ ,\label{metric}
\ee
obtaining
\be
{\cal S}=\int dt\, a^3\left[-3\frac{H^2}{\kappa^2 N}+\frac{1}{2}\frac{\dot\Phi^2}{N}+\frac{3}{2}\frac{H^2w^2}{N^3}\dot\Phi^2-N\frac{\lambda}{4}\Phi^4\right]\ ,\label{t2}
\ee
where $H=\dot a/a$ and $(\dot{})=d/dt$.

The only independent gravity equation, with FRW symmetries, is then recovered by considering the Hamiltonian constraint obtained by varying the action (\ref{t2}) with respect to the lapse $N$ and then setting it to $1$ by time reparameterization invariance \cite{wald}. The field equation for $\Phi$ corresponds instead to the variation of (\ref{t2}) with respect to $\Phi$.

The Hamiltonian constraint and field equation are
\be
H^2=\frac{\kappa^2}{6}\left[\dot\Phi^2\left(1+9H^2w^2\right)+\frac{\lambda}{2}\Phi^4\right]\ ,\cr
\partial_t\left[a^3\dot\Phi\left(1+3H^2w^2\right)\right]=-a^3\lambda\Phi^3\ .\label{background}
\ee
We will ask the solution to obey the following inequalities
\be
H\gg \frac{1}{3w}\ , ~~~ 9H^2w^2\dot\Phi^2\ll \frac{\lambda}{2}\Phi^4\ , ~~~ -\frac{\dot H}{H^2}\ll1\ ,\label{SR2}
\ee
where the last two are the usual slow roll conditions. Of course (\ref{SR2}) must be cross checked afterwords.

With (\ref{SR2}) we find
\be
H^2\simeq \frac{\kappa^2}{12}\lambda\Phi^4\ ,\label{Hour}
\ee
and
\be
\ddot\Phi+3H\dot\Phi=-4/(w^2 \kappa^2\Phi)\ .
\ee
By considering the extra slow roll condition
\be
|\ddot\Phi|\ll3H|\dot\Phi|\ ,\label{extra}
\ee
we finally get
\be
\dot\Phi\simeq-\frac{4}{3Hw^2\kappa^2\Phi}\ .\label{dotp}
\ee
The quantum gravity constraint $R\simeq 12H^2\ll 1/(2\kappa^2)$ implies
\be
\Phi^4\ll \frac{1}{2\kappa^4\lambda}\ .\label{max}
\ee
We now need to cross check the various constraints (\ref{SR2},\ref{extra}).
As $\lambda<1$, we have the stronger constraint
\be
\Phi^6\gg 32/(w^2 \kappa^4\lambda)\ ,\label{Pgg}
\ee
Combining this with (\ref{max}), we have
\be
w/\kappa\gg 10\times \lambda^{1/6}\ .\label{w/k}
\ee
\section{Cosmological Perturbations}

Following \cite{malda} we will study the fluctuations around the slow-roll inflationary solution in the ADM formalism \cite{wald}. In this formalism a general metric can be decomposed as
\be
ds^2=-N^2dt^2+h_{ij}\left(dx^i-N^i dt\right)\left(dx^j-N^j dt\right)\ ,
\ee
and all geometrical quantities are described by defining a spatial covariant derivative $D_i$ a three-dimensional
curvature ${}^{(3)}R$ (both constructed on $h^{ij}$) and finally the extrinsic curvature
$K_{ij}=\frac{1}{2N}\left(\dot{h}_{ij}-D_i N_j-D_j N_i\right)$.

In \cite{higgsinf} it has been proved that the theory (\ref{theory}) only propagates three degrees of freedom.
Two of those come from
gravity and one from the scalar, as it happens for minimally coupled gravity to a scalar field. Moreover, since a scalar field in a FRW background is rotational invariant, the decoupling of scalar perturbations with tensorial is granted \cite{liddle}.

The action (\ref{theory}) is invariant under the diffeomorphisms group, as a result of reparametrization invariance.
To simplify the calculations we can then impose the gauge \cite{malda}
\be
\delta \Phi=0\, , ~~~h_{ij}=a(t)^2\Big{[}(1+2 \zeta)\delta_{ij}+\gamma_{ij}\Big{]}\, , ~~~ ~~~
\partial^i\gamma_{ij}={\gamma^i}_{i}=0\ ,
\ee
where $\zeta$ parametrizes the scalar perturbations and $\gamma_{ij}$ the gravitational waves.

The expanded second order action coming from (\ref{theory}) is then
\be
\!\!\!\!\!\!\!\!\!\!\!\!S=\frac{1}{2\kappa^2}\int \!\!d^3x dt\sqrt{h}N\left(\!\!{}^{(3)}\!R(1\!+\!\frac{w^2\kappa^2\dot{\Phi}^2}{2N^2})\!+\!
(K_{ij}K^{ij}\!-\!K^2)(1\!-\!\frac{w^2\kappa^2\dot{\Phi}^2}{2N^2})\!+\!
\frac{\kappa^2\dot{\Phi}^2}{N^2}\!-\!
2\kappa^2V\right)\ , \label{ac}
\ee
where $V=\lambda\Phi^4/4$ is the scalar potential.

The Hamiltonian (variation with respect to $N$) and momentum (variations with respect to $N^i$)
constraints are found to be
\be
D_i\left[(1-\frac{w^2\kappa^2\dot{\Phi}^2}{2N^2})(K^{ij}-h^{ij}K)\right]=0\,, \label{D}
\ee
\be
(1-\frac{w^2\kappa^2\dot{\Phi}^2}{2N^2}){}^{(3)}\!R-(1-\frac{3w^2\kappa^2\dot{\Phi}^2}{2N^2})(K_{ij}K^{ij}-K^2)-
\frac{\kappa^2\dot{\Phi}^2}{N^2}-
2\kappa^2V=0\ .\label{N}
\ee

\subsection{Scalar Perturbations}

Considering a perturbative solution of the form $N=1+N_1$ and $N^i=\partial_i \psi+ N^i_T$, we find, by using (\ref{D}) and (\ref{N}),
\be \label{NN}
\!\!\!\!\!\!\!\!\!\!\!\!N_1=\frac{ \dot{\zeta}}{H}\,,~~ \psi=-\frac{2-w^2\kappa^2\dot{\Phi}^2}{2-3w^2\kappa^2\dot{\Phi}^2}
\frac{\dot{\zeta}}{a^2H}+\chi\, ,~~
\partial^2\chi=-\kappa^2\frac{1+9w^2H^2}{2-3w^2\kappa^2\dot{\Phi}^2}\frac{\dot{\Phi}^2}{H^2}\dot{\zeta}\, ,~~ \nabla_i N^i_T=0\ .
\ee
We may now use (\ref{NN}) into the action (\ref{ac}) and expand it up to second order in $\zeta$. The result is
\be
S_\zeta&=&\frac{1}{2\kappa^2}\int d^3x dt\left\{e^{\rho+\zeta}(1+\frac{\dot{\zeta}}{H})\left[\left(1+\frac{w^2\kappa^2
\dot{\Phi}^2}{2(1+
\frac{\dot{\zeta}}{H})^2}\right)
[-4\partial^2\zeta-2(\partial\zeta)^2]-2\kappa^2Ve^{2(\rho+\zeta)}]+\right.\right.\nonumber \\
&+&e^{3(\rho+\zeta)}\frac{1}{1+\frac{\dot{\zeta}}{H}}
\left[-6(H+\dot{\zeta})^2\left(1-\frac{w^2\kappa^2\dot{\Phi}^2}{2(1+
\frac{\dot{\zeta}}{H})^2}\right)+\kappa^2\dot{\Phi}^2\right]-\\\nonumber &-&
2w^2\kappa^2 (\partial\zeta)^2\, \partial_t\!\left(\frac{\dot{\Phi}^2}{H}\frac{2-w^2\kappa^2\dot{\Phi}^2}{2-3w^2\kappa^2
\dot{\Phi}^2}
e^{\rho+\zeta}\right)+
\left.4w^2\kappa^4e^{3(\rho+\zeta)}\frac{1+9w^2H^2}{2-3w^2\kappa^2\dot{\Phi}^2}\frac{\dot{\Phi}^4}{H^2}\dot{\zeta}^2\right\}\ ,
\ee
where, following \cite{malda}, we redefined $a(t)=e^{\rho(t)}$ so that $H=\dot{\rho}$.

Using the background equations (\ref{background}) together with the conditions $w^2\kappa^2\dot{\Phi}^2\ll1$
(Eq. (\ref{SR2})) and $w^2H^2\gg1$, we obtain the following quadratic action in $\zeta$
\be
S_\zeta=\frac{1}{2}\int d^3x dt \left\{3w^2\kappa^2\dot{\Phi}^2\left[e^{3\rho} (1+6w^2\kappa^2\dot{\Phi}^2)\dot{\zeta}^2
-e^{\rho}(1-\frac{13}{3}w^2\kappa^2\dot{\Phi}^2)
(\partial \zeta)^2\right]\right\}\ .
\ee
In order to quantize the system, we  need to canonically normalize the kinetic term.
Considering that, during slow roll, corrections proportional to $\ddot\Phi$ are sub-leading,
we can define the canonically normalized variable
\be
\tilde \zeta=\sqrt{3}w\kappa\dot\Phi\sqrt{1+6w^2\kappa^2\dot\Phi^2}\, \, \, \zeta\ .
\ee
Then, the quadratic action may be written as
\be
S_\zeta=\frac{1}{2}\int d^3x dt e^{3\rho}\left[ \dot{\tilde\zeta}^2-e^{-2\rho}c_s^2(\partial \tilde\zeta)^2\right]\ ,\label{can}
\ee
where the sound speed of the scalar perturbations is
\be
0<c_s^2=\frac{3-13w^2\kappa^2\dot{\Phi}^2}{3+18w^2\kappa^2\dot{\Phi}^2}<1\ .
\ee
The power spectrum  of scalar perturbations is defined as the two point correlation function of $\zeta$
 at the horizon crossing $c_s k=aH$, where $k$ is the Fourier mode in the action (\ref{can}) or, more precisely,
\be
\langle \zeta_{\bf{k}}\zeta_{\bf{k'}}\rangle\Big{|}_{c_sk=aH}=\frac{2\pi^2}{k^3}{\cal{P}}_S\, \delta^{(3)}(\bf{k+k'})\ ,
\ee
where
\be
\zeta=\int \frac{d^3 k}{(2\pi)^{3/2}} \zeta_{\bf{k}}(t)e^{i\bf{k\cdot x }}\ .
\ee
By standard canonical quantization of the perturbation $\tilde\zeta$ during inflation \cite{liddle}, one can then easily find
\be
{\cal{P}}_S=\frac{\kappa^2 H^2}{4\pi^2} \frac{1}{12w^2\kappa^2\dot{\Phi}^2}(1+\frac{19}{2}w^2\kappa^2\dot{\Phi}^2)\Big{|}_{aH=c_s k}\ .
\ee
Defining the slow-roll parameters \cite{liddle}
 \be
 \epsilon=-\frac{\dot{H}}{H^2}=\frac{3}{2} \kappa^2 w^2\dot{\Phi}^2\, , ~~~\eta_H=-
 \frac{\ddot{\Phi}}{H \dot{\Phi}}=-\frac{9}{4}\kappa^2
 w^2\dot{\Phi}^2=-\frac{3}{2}\, \epsilon\ ,\label{slowroll}
\ee
where (\ref{Hour},\ref{dotp}) has been used,
we finally find that
\be
{\cal{P}}_S=\frac{\kappa^2 H^2}{4\pi^2} \frac{1}{8\epsilon}\ ,\label{PS}
\ee
to leading order in $\epsilon$.

The spectral index of scalar perturbation is given by \cite{liddle}
\be
n_S-1=\frac{d\ln {\cal{P}}_S}{d\ln k}=2 \frac{d\ln H}{d\ln k}-\frac{d \ln \epsilon}{d\ln k}\ .
\ee
By using (\ref{slowroll}) we find
\be
n_S=1-2\epsilon+2\eta_H=1-5\epsilon\ .
\ee
The running of the spectral index can also be specified to be
\be
\alpha=\frac{dn_S}{d\ln k}=-15 \epsilon^2\ ,
\ee
where the definitions (\ref{slowroll}) have been used.
Taking for $n_S$ the maximum likelihood value as specified by the 7-years WMAP observations of the
Cosmic Microwave Background Radiation (CMBR) \cite{wmap}, {\it i.e.} $n_S=0.969$, we get
\be
\epsilon=0.0062\ .\label{epsilonvalue}
\ee

\subsection{Gravitational Waves}

Let us now consider tensor perturbations for the action (\ref{theory}).
In the ADM formalism, a convenient gauge to analyze tensor perturbations is
\be
h_{ij}=a(t)^2(\delta_{ij}+\kappa \gamma_{ij})\, , ~~~\partial^i\gamma_{ij}={\gamma^i}_{i}=0\ .
\ee
After inserting this into the action (\ref{theory}) and keeping only quadratic terms in $\gamma_{ij}$, we get
the quadratic action for gravitational waves
\be
S_g=\int d^3x dt \frac{1}{8} \left[(1-\frac{1}{2}w^2\kappa^2\dot{\Phi}^2)\, a^3\, \dot{\gamma}_{ij}\dot{\gamma}_{ij}-
(1+\frac{1}{2}w^2\kappa^2\dot{\Phi}^2)
\, a\, (\partial_k \gamma_{ij})^2\right]\ .
\ee
We can now canonically normalize the graviton as
\be
\tilde \gamma_{ij}=\sqrt{\frac{1-\frac{1}{2}w^2\kappa^2\dot\Phi^2}{4}}\gamma_{ij}\ ,
\ee
as no ghosts are propagated thanks to $\frac{1}{2}w^2\kappa^2\dot\Phi^2\ll1$. In the new variable we have
\be
S_g=\int d^3x dt\ a^3\frac{1}{2}\left[\dot{\tilde \gamma}_{ij}\dot{\tilde \gamma}_{ij}-c_g^2\, a^{-2}\, (\partial_k \tilde\gamma_{ij})^2\right]\ ,\label{grav}
\ee
where we defined the sound speed
\be
c_g^2=\frac{2+w^2\kappa^2\dot{\Phi}^2}{2-w^2\kappa^2\dot{\Phi}^2}\simeq 1+w^2\kappa^2\dot{\Phi}^2>1\ .
\ee
Gravitational waves then propagate with superluminal velocity during Inflation.

Tensor perturbation may be expanded in Fourier modes
\be
\gamma_{ij}=\int \frac{d^3 k}{(2\pi)^{3/2}} \gamma_{\bf{k}}(t) e_{ij}(\bf{k}) e^{i\bf{k\cdot x }}\ ,
\ee
where $e_{ij}(\bf{k})$ is the polarization tensor.
The power spectrum of gravitational waves is then defined as \cite{liddle}
\be
\langle \gamma_{\bf{k}}\gamma_{\bf{k'}}\rangle\Big{|}_{aH=c_g k}=\frac{2\pi^2}{k^3}{\cal{P}}_T\, \delta^{(3)}(\bf{k+k'})\ ,
\ee
where the correlator is easily found by canonically quantizing the action (\ref{grav}). Explicitly we have
\be
{\cal{P}}_T= \frac{\kappa^2 H^2}{4\pi^2 \Omega^2 c_g^3}\Big{|}_{aH=c_g k}\ ,
\ee
where
\be
\Omega^2=1-\frac{1}{2}w^2\kappa^2\dot{\Phi}^2\ .
\ee

To leading order in $w^2\kappa^2\dot{\Phi}^2$, we have $\Omega\approx 1\, , ~c_s\approx 1$. Therefore, the spectrum of gravitational waves for the New Higgs Inflation
is approximately the same of the chaotic inflationary one.

We can finally define the tensor-to-scalar ratio $r={\cal{P}}_T/{\cal{P}}_S$. To leading order in slow roll we have
\be
r=12 w^2\kappa^2\dot{\Phi}^2\Big{|}_{aH=c_g k}=8\epsilon\, .
\ee
\section{Extracting the Parameters}

Scalar perturbations in the CMBR are very accurately measures by \cite{wmap}. The measured Power spectrum has an amplitude
\be
P_S=2.38\times 10^{-9}\ ,\label{P}
\ee
at the pivot scale $k=0.002\ \rm{Mpc}^{-1}$. With this data and the value of $\epsilon$ (\ref{epsilonvalue}),
we are now able to fix the parameters of the New Higgs inflationary theory (\ref{theory}).

The first information one could obtain from the measure (\ref{P}), by using (\ref{PS}), is that the curvature value during Inflation is
\be
R\simeq 12 H^2\simeq 5.6\times 10^{-8} M_p^2\ ,\label{MH}
\ee
where we used the reduced Planck mass $M_p=\kappa^{-1}=2.4\times 10^{18}\ \rm{GeV}$. It is then clear that the curvature during inflation is much below the Planck scales. We now need to fulfill the inequality (\ref{w/k}).

The Higgs value during Inflation ($\Phi_0$) may be read by using (\ref{Hour},\ref{MH}) to be                                                                        
\be                                                                                                                                                                       
\Phi_0=0.0154\, \,  \lambda^{-1/4} \,\, M_p\ .                                                                                                                             
\ee                                                                                                                                                                       
Using also the (\ref{slowroll}), and (\ref{MH}) together with the value (\ref{epsilonvalue}), we obtain the value of the dimensionfull parameter $w$ to be                                                                          
\be                                                                                                                                                                       
w^{-1}=5.08 \times 10^{-8} \lambda^{-1/4}\,\, M_p=3.3 \times 10^{-6}\, \, \Phi_0\ .                                                                                         
\ee                                                                                                                                                                       
Clearly this value of $w$ satisfies the requirement (\ref{w/k}) for $\lambda\gg 10^{-29}$! 
                                                                         
Although the exact value of $\lambda$ is not known during inflation, one may use the values of $\lambda$ at electroweak scale, which, by direct Higgs boson searches as well as from global fit to                                               
electroweak precision data, is restricted to be in the range $0.11<\lambda\lesssim 0.27$ \cite{bound}.
This gives values for $\Phi_0$ and $w$ in the range 
\be                                                                                         
&&2.1\times 10^{-2}\ M_p<\Phi_0<2.7\times 10^{-2}\ M_p\, , \nonumber  \\                                                                                                  
&& 7\times 10^{-8}\ M_p<w^{-1}<8.8\times 10^{-8}\ M_p\ \, ,\label{running}
\ee
showing that the approximation $\Phi_0\gg v$ is consistent.

Due to the running of $\lambda$, these values should not however be taken as precise measures. Nevertheless, we expect the orders of magnitude in (\ref{running}) to still work during the inflationary scales. In fact, if for example we consider the running of $\lambda$ in a Minkowskian background \cite{djouadi} at the inflationary scale $\sim 10^{14} {\rm GeV}$, we find that (\ref{running}) will only change by a numerical factor of order $1$. To be more precise, let us consider the lower value $\lambda_{l}\approx 0.186$ at electroweak scale, which saturates the stability bound \cite{djouadi}. This runs to the value $\lambda_l\approx 0.019$ at $(10^{14} {\rm GeV})$ and would correct the magnitude of $\Phi_0$ and $w$ by just a factor of $2$. As a result, we may safely conclude that $\Phi_0\sim 10^{16} {\rm GeV}$ and $w^{-1}\sim 10^{9}-10^{10} {\rm GeV}$.    

 Finally we can calculate the number of e-foldings of Inflation. By definition, the number of e-foldings $N$ of Inflation is \cite{liddle}
 \be
N=\int^{t_f}_{t_i} dt H=\int^{\Phi(t_f)}_{\Phi(t_i)}\frac{H}{\dot\Phi}d\Phi\ .
 \ee
Considering that $\Phi(t_f)\ll\Phi(t_i)$ we finally have
\be
N\simeq \frac{1}{3\epsilon}\simeq 54\ ,
\ee
where (\ref{epsilonvalue}) and (\ref{P}) has been used. Note that $54$ e-foldings are enough to solve the
cosmological problems \cite{liddle}.

Vice-versa, one could impose the number of e-foldings $N$ in relation to a particular re-heating temperature of the Universe \cite{liddle} and predict the spectral index to be
\be
n_S=1-\frac{5}{3 N}\ .
\ee
For example for $N= 60$ we would have $n_S=0.972$, which is within the errors of the present data \cite{wmap}.

\section{Conclusions}
In the New Higgs Inflation the inflationary phase of the early Universe is
triggered by the Standard Model Higgs boson, whose kinetic term is non-minimally coupled to the Einstein tensor. As shown in \cite{higgsinf}, this non-minimal coupling allows a primordial inflating background of the Universe within the Standard Model Higgs self coupling value. Without this non-minimal coupling, the quantum gravity bound would require the Higgs
self-coupling $\lambda$ to be orders of magnitude lower than its collider experimental lower bound \cite{chaotic}. Whereas, other previous attempts to non-minimally couple the Higgs to gravity in order to reproduce successful inflation (see \cite{sha}), seems to fail due to a Higgs strong coupling to gravity during Inflation \cite{dub}.

The non-minimal coupling introduced in \cite{higgsinf} propagates, in an inflating background, only three degrees of freedom: a scalar degree of freedom associated to the Higgs boson and two degrees of freedom associated to the primordial gravitational waves.

In this paper we therefore studied primordial scalar and tensorial perturbations produced during the inflationary phase driven by the New Higgs Inflaton. We found that the observed CMBR spectrum
of scalar perturbations, as measured during the first 7 years of WMAP \cite{wmap}, is compatible with the New Higgs Inflation. Tensor perturbations instead, although being compatible with experimental data, propagate at superluminal speed and the ratio of scalar-to-tensor perturbations is suppressed by the
slow-roll parameter $\epsilon$.

Finally, by matching the predictions of our model with observations we estimated the value of the dimensionfull parameter $w$
coupling the kinetic term of the Higgs to gravity. We found that its value is around the geometric mean of Electroweak and Planck scale. Related to that, we calculated the curvature during the
New Higgs Inflation and showed that, in compatibility with the semiclassical approximation for gravity, it is well below the Planck scales.

 \section*{Acknowledgments}
CG thanks Fedor Bezrukov for useful discussions on the predictability of the theory. CG is sponsored by the Humboldt Foundation. AK would like to thank Nick Tracas for discussions on the running of $\lambda$. This work is partially supported by the European Research and Training Network MRTPN-CT-2006 035863-1 and the PEVE-NTUA-2009 program.

\end{document}